\documentstyle[emulateapj,psfig]{article}
\def\gs{\mathrel{\raise0.35ex\hbox{$\scriptstyle >$}\kern-0.6em 
\lower0.40ex\hbox{{$\scriptstyle \sim$}}}}
\def\ls{\mathrel{\raise0.35ex\hbox{$\scriptstyle <$}\kern-0.6em 
\lower0.40ex\hbox{{$\scriptstyle \sim$}}}}
\def\Mdot{\mathrel{\kern0.4em\raise1.6ex\hbox{$\bf .$}\kern-0.75em 
\lower0.10ex\hbox{{$M$}}}\kern0.2em}
\submitted{Received 1999 March 5; accepted: 1999 -- --}

\lefthead{Smail et al.}
\righthead{Faint Radio Sources in the core of A\,851}

\begin{document}

\title{{\it Hubble Space Telescope} Near-infrared and Optical Imaging of Faint Radio Sources in the Distant Cluster Cl\,0939+4713}

\author{Ian Smail,$\!$\altaffilmark{1}
G.\ Morrison,$\!$\altaffilmark{2} M.\,E.\ Gray,$\!$\altaffilmark{3}
F.\,N.\ Owen,$\!$\altaffilmark{2} R.\,J.\ Ivison,$\!$\altaffilmark{4}
J.-P.\ Kneib\altaffilmark{5}  \& R.\,S.\ Ellis\altaffilmark{3} }

\affil{\tiny 1) Department of Physics, University of Durham, South Road, 
Durham DH1 3LE, UK}
\affil{\tiny 2) NRAO, P.O.\ Box 0, 1003 Lopezville Road, Socorro, NM 87801}
\affil{\tiny 3) Institute of Astronomy, Madingley Road, Cambridge
CB3 OHA, UK}
\affil{\tiny 4) Department of Physics and Astronomy, University College
London, Gower Street, London WC1E 6BT, UK}
\affil{\tiny 5) Observatoire Midi-Pyr\'en\'ees, CNRS-UMR5572,
14 Avenue E.\ Belin, 31400 Toulouse, France}

\setcounter{footnote}{6}

\begin{abstract}
We present deep {\it Hubble Space Telescope} {\it NICMOS} near-infrared
and {\it WFPC2} optical imaging  of a small region in the core of
the distant rich cluster Cl\,0939+4713 ($z=0.41$). We compare the
optical and near-infrared morphologies of cluster members and find
apparent small-scale optical structures within the galaxies which are
absent in the near-infrared.  We conclude that strong dust obscuration
is a common feature in the late-type galaxies in distant clusters.
We then concentrate on a sample of ten faint radio galaxies lying
within our {\it NICMOS} field and selected from a very deep 1.4-GHz
VLA map of the cluster with a 1\,$\sigma$ flux limit of 9\,$\mu$Jy.
Using published data we focus on the spectral properties of the eight
radio-selected cluster members and show that these comprise a large
fraction of the post-starburst population in the cluster.  The simplest
interpretation of the radio emission from these galaxies is that they are
currently forming massive stars, contradicting their classification as
post-starburst systems based on the optical spectra. We suggest that this
star formation is hidden from view in the optical by the same obscuring
dust which is apparent in our comparison on the optical and near-infrared
morphologies of these galaxies.  We caution that even in the restframe
optical the effects of dust cannot be ignored when comparing samples
of distant galaxies to low-redshift systems, particularly if dust is as
prevelant in distant galaxies as appears to be the case in our study.  
\end{abstract}


\section{Introduction}

The characteristics of the galaxy populations in the cores of rich
clusters appear to vary strongly as a function of redshift, with
an increasing fraction of blue galaxies in clusters at $z>0.2$ (the
Butcher-Oemler effect, Butcher \& Oemler 1984).  The search for the
processes responsible for these rapid changes has become a vigorous field
of research and at least for the blue cluster galaxies a broad concensus
has been reached about the nature of this population.  The bulk are
star-forming cluster members, they cover a wide range of luminosities,
from a few $L^\ast$ downwards, and have emission line strengths which
indicate moderate to high star formation rates (Abraham et al.\ 1996;
Balogh et al.\ 1999; Barger et al.\ 1996; Couch \& Sharples 1987; Dressler
\& Gunn 1992; Dressler et al.\ 1999, D99; Fabricant et al.\ 1994; Fisher
et al.\ 1998; Lavery \& Henry 1988).  The optical morphologies of these
galaxies were investigated firstly with high-resolution ground-based
(Lavery et al.\ 1992) and more recently with {\it Hubble Space Telescope}
({\it HST}\,) imaging (Dressler et al.\ 1994a; Couch et al.\ 1994, 1998;
Oemler et al.\ 1997; Smail et al.\ 1997, S97; Fabricant et al.\ 1999).
These studies have shown that most of the blue star-forming galaxies
have strong disk components and a large fraction also appear to be
disturbed, with structures suggesting  mergers or tidal interactions.
Thus the increasing proportion of blue galaxies within distant clusters
is associated with an increasing number of actively star-forming
disk galaxies in these environments.  The large fraction of apparently
disturbed galaxies suggests that galaxy-galaxy interactions or the effects
of the cluster tidal field on these disk galaxies could be responsible
for the Butcher-Oemler effect (Moore, Lake \& Katz 1998).

The spectroscopic studies of distant clusters  uncovered another
population of luminous galaxies which is not seen in similar numbers in
local clusters.  These are galaxies with strong Balmer absorption
(typically measured using the H$\delta$\,$\lambda$4104\AA\ line and
coming predominantly from A stars) and no  [O{\sc
ii}]\,$\lambda$3727\AA\ emission.  The large population of A stars is a
sign that the galaxy was actively forming massive stars in the recent
past ($\ls 1$\,Gyr), while the lack of emission lines suggests that
this star formation has now ceased.  Together the spectral properties
are interpreted as showing that the galaxy is in a post-starburst
phase.  Almost 20\% of the luminous galaxies within the cores of
distant clusters fall in this post-starburst class (e.g.\ D99, although
see Balogh et al.\ 1999) and it has been suggested that this high
fraction indicates that almost all cluster galaxies pass through this
phase (Barger et al.\ 1996).

There is an increasing realisation in the community of the importance
of dust obscuration in defining the apparent properties of galaxies in
the near-UV and optical, particularly at high redshifts.  This issue
has been highlighted by the disagreement over the form of the star
formation history of the Universe as estimated from the variation in
the ultraviolet luminosity density with redshift (e.g.\ Lilly et
al.\ 1996; Steidel et al.\ 1999)  compared to star formation indicators
at longer wavelengths such as H$\alpha$ emission (Yan et al.\ 1999) or
reradiated star light detected in the sub-millimeter (e.g.\ Blain et
al.\ 1999).  The longer wavelength tracers tend to find higher star
formation densities, a result which has been attributed to dust
absorption in the UV and optical (Calzetti \& Heckman 1999; Meurer et
al.\ 1997; Pettini et al.\ 1998).

The issue of the effects of dust on the evolutionary cycle
associated with the optical Butcher-Oemler effect has been disregarded by
most investigations to date. However, the recent analysis of the spectral
catalog for the MORPHS sample (D99) by Poggianti et al.\ (1999, P99) has
uncovered evidence for dust obscuration in at least one spectral class  --
e(a) galaxies.  These galaxies have detectable [O{\sc ii}] emission and
relatively strong Balmer absorption (EW(H$\delta)\geq 4$).  These spectral
features, along with other properties of the galaxies and of similar
spectrally-classified local systems (P99; Poggianti \& Wu 1999), were
interpreted by P99 as probably arising from a dust-obscured starburst,
with the [O{\sc ii}] line strength suppressed by dust absorption.
The e(a) class is naturally identified as the progenitor of the large
population of post-starburst galaxies in the clusters.

To obtain a complete view of the star formation properties of cluster
galaxies free from the effects of dust obscuration we need to
complement the optical studies with observations at longer wavelengths,
from the near-infrared to the sub-mm and radio.  In particular,
observations in the sub-mm and far-infrared can trace the amount of
star light absorbed by dust and reradiated at longer wavelengths.  For
the most extreme starburst galaxies in the local Universe this
reradiated emission dominates the bolometric luminosity of the galaxies
(e.g.\ Sanders \& Mirabel 1996).  Unfortunately, current sub-mm and
far-infrared surveys lack the sensitivity to study all but the most
extreme starburst galaxies ($\Mdot \gs 100 M_\odot$ yr$^{-1}$) at
moderate and high redshift.  However, at radio wavelengths, the
synchotron radiation from electrons accelerated by supernovae leads to
a tight correlation between the radio and far-infrared fluxes of
star-forming galaxies across a wide range in luminosity (Condon 1992).
This means that in the absence of a radio-loud active nucleus the radio
flux of a star-forming galaxy can be used to estimate its current
massive star formation rate.  

By employing the huge collecting area of the VLA  we can undertake a
sensitive radio survey for star formation in obscured galaxies down to limits
of $\Mdot \sim 5$--$10 M_\odot$ yr$^{-1}$ at $z\sim 0.5$.  To this end
the VLA has been used in an ambitious survey to determine the evolution
of the radio populations in a sample of very-rich clusters out to
$z=0.4$ (Morrison et al.\ 1999).   The survey contains 14 clusters at
$z=0.10$--0.18, 12 more between $z=0.20$--0.25 and a further four at
higher redshifts, $z=0.38$--0.41 (the radio observations discussed in
this paper are of the most distant cluster in this survey).  Morrison et
al.\ (1999) find that the population of radio galaxies with luminosities
above $2 \times 10^{22}$\,W\,Hz$^{-1}$ in the most extreme clusters has
increased by a factor of $\sim 5$--6 out to $z=0.4$.  They conclude
that the radio galaxy population in very rich clusters has evolved
significantly over the last 5\,Gyrs.

In this paper we report on high resolution near-infrared imaging from
{\it HST} and sensitive radio observations with the VLA  of a small field
in the core of the distant cluster Cl\,0939+4713 (A\,851, $z=0.41$) for
which high-resolution optical {\it HST} images and extensive archival
spectroscopic observations are available.  We combine these datasets
to study the characteristics of galaxies in distant clusters and to search
for the signatures of dust obscuration on their apparent properties.

In \S2 we describe the {\it HST} optical and near-infrared imaging as well
as the deep VLA 1.4-GHz map of this field.    In \S3 we then present
our results on the properties of the radio-selected galaxies within our
{\it HST} field drawing on published spectroscopy.  We discuss these in
\S4 and give our main conclusions in \S5.    Throughout this paper we
use $H_\circ = 50$\,km\,s$^{-1}$\,Mpc$^{-1}$ and a $\Omega_\circ=1$,
$\Lambda=0$ cosmology.  In this geometry $1''$ corresponds to 6.5\,kpc
at $z=0.41$.

\section{Observations}

\subsection{{\it NICMOS} Imaging}

The near-infrared {\it NICMOS} observations of Cl\,0939+4713 discussed
here were obtained with Camera~3 during the special campaign in January
1998.  Due to the distortion of the {\it NICMOS} dewar the wide-field
Camera~3 (field-of-view 51.2$''$ square, sampling 0.20$''$ per pixel)
was not par-focal with the other instruments onboard {\it HST} and
to allow the diffraction-limited operation of this camera  a special
campaign was organised during which the main telescope  optics
were refocused on the Camera~3 focal plane.

The observations comprise a mosaic of four pointings through the F160W
filter covering a combined field of around $100'' \times 100''$.  Each
pointing consists of four individual exposures of 1215\,s, each
spatially offset by 1.1$''$ on a square grid to remove bad pixels and
reduce the effects of flatfield variations.  The total integration time
per pixel for the mosaic is thus  4.86\,ks.  The data were reduced in a
standard manner using the {\it NICMOS} data pipeline ({\sc calnica})
and the most recent calibration files taken from the archive.  Any
remaining features in the background were removed with the {\sc unpedestal}
software kindly provided by Dr.\ R.\ van der Marel which worked well,
and the final mosaicing was achieved using {\sc imcombine} in {\sc
IRAF}.  The mosaic was photometrically calibrated to give magnitudes
on the $H_{160}$ Vega system.  This image provides restframe
$J$-band morphologies at a resolution of around 1.3\,kpc for cluster
members (Fig.~1).  

To create a complete catalog of objects detected within the {\it NICMOS}
mosaic we analysed the frame using the SExtractor image analysis package
(Bertin \& Arnoults 1996).  We adopted detection criteria of a minimum
area of  $\geq 0.4$\,sq.\ arcsec (10 pixels) above an $\mu_{160} =
24.0$ mag arcsec$^{-2}$ isophote and using these we detect 303 sources
brighter than the approximate 3\,$\sigma$ point-source flux limit of
$H_{160}=23.5$.

\subsection{Radio Data}

The VLA observations of Cl\,0939+4713 were obtained in B configuration
at 1.4\,GHz between 1996 January 06--08.  The total integration time of
the map is 60.7\,ks at an effective resolution of 5.0$''$ (33\,kpc at
$z=0.41$). The map was cleaned and analysed using {\sc aips}. Details
of this complex procedure are given in Morrison et al.\ (1999).  The map
reaches a 1\,$\sigma$ noise level of 9\,$\mu$Jy\,beam$^{-1}$.  We detect
10 sources above 27\,$\mu$Jy (3\,$\sigma$) within the field defined by
the overlap region of our {\it NICMOS} F160W mosaic and {\it Wide Field
and Planetary Camera 2} ({\it WFPC2}\,)  F702W exposures (see Fig.~1).
All ten sources have bright optical/near-infrared counterparts in our
{\it HST} images.  The properties of these ten sources are discussed
in \S3.  The positions, fluxes and restframe radio power (using the
redshifts from Table~2 and assuming an $S\propto \nu^{-0.7}$ spectrum)
are reported in Table~1.  Unfortunately due to the modest resolution
of the map only two sources, \#122 and \#296, appear to be resolved
and then only marginally.  The radio and optical/near-IR
images are aligned to an rms accuracy of $0.4''$ using the 8 brightest
radio sources in the {\it NICMOS} field.

The flux limit of the radio sample corresponds to a radio power of just
$2\times 10^{22}$\,W\,Hz$^{-1}$ at the redshift of the cluster.
Comparing this limit to those achieved by radio surveys of local
clusters we note that Gavazzi \& Boselli (1999) detect galaxies as
faint as $3 \times 10^{20}$\,W\,Hz$^{-1}$ in their analysis of the
radio luminosity function in the Virgo cluster (a factor of roughly
$100\times$ closer to us than Cl\,0939+4713) using the NVSS 1.4\,GHz
survey (Condon et al.\ 1998).
  
%
%
\begin{table*}[hbt]
{\scriptsize
\begin{center}
\centerline{\sc Table 1}
\vspace{0.1cm}
\centerline{\sc Properties of the Radio Galaxy Sample}
\vspace{0.3cm}
\begin{tabular}{cccccl}
\hline\hline
\noalign{\smallskip}
 {Source} & {R.A.} & {Dec.} & {$S_{1.4}$} & {$P_{1.4}$} & {Comments} \cr
 ID & \multispan2{ ~(J2000) } & ($\mu$Jy) & ($10^{22}$\,W\,Hz$^{-1}$) & \cr
\hline
\noalign{\smallskip}
~36 & 09 42 58.16 & +46 59 53.4 & ~$79\pm 9$  & ~5.9  & \cr
122 & 09 42 57.62 & +46 59 45.3 & $204\pm 9$  & 15.2  & Resolved?\cr
224 & 09 42 55.80 & +46 59 40.4 & ~$55\pm 9$  & ~4.0  & \cr
230 & 09 43 02.32 & +46 59 28.7 & ~$52\pm 9$  & ~3.8  & $P_{1.4}$ assumes $z=0.41$\cr
296 & 09 43 02.81 & +46 59 24.2 & $644\pm 9$  & 47.6  & Resolved?\cr
399 & 09 42 57.84 & +46 59 13.3 & ~$43\pm 9$  & ~3.1  & \cr
426 & 09 42 56.21 & +46 59 12.0 & ~$32\pm 9$  & ~2.4  & \cr
536 & 09 43 00.03 & +46 58 53.4 & $196\pm 9$  & 14.5  & \cr
610 & 09 42 57.34 & +46 58 50.4 & $158\pm 9$  & 11.7  & \cr
640 & 09 42 55.08 & +46 58 41.7 & ~$57\pm 9$  & ~2.6  & \cr
\noalign{\hrule}
\noalign{\smallskip}
\end{tabular}
\end{center}
}
\end{table*}

\subsection{{\it WFPC2} Imaging}

These observations were obtained as part of the Early Release
Observations during the Science Verification Phase after the
installation of the {\it WFPC2} on board {\it HST}.  The observations
comprise a total of 21.0\,ks of integration through the F702W filter
(Dressler et al.\ 1994b).  The exposures were dithered by integer
pixels to allow the removal of defects and cosmic ray events.  The
position of the corrective optics within {\it WFPC2}, to correct for
the aberration of the primary mirror, was not optimal during these
observations and combined with a higher than ideal operating
temperature for the CCDs these images are not as cosmetically clean as
later exposures with {\it WFPC2}.  Nevertheless, they provide deep
($R\sim 26$) imaging of the core of the cluster at 0.1--0.2$''$
resolution in the restframe $B$-band.
  
The reduction and analysis of the F702W exposure of this field is
described in more detail in Smail et al.\ (1997), who provide a
complete catalog of objects detected in the field with photometry
calibrated to the $R_{702}$ passband described by Holtzmann et
al.\ (1995).  Smail et al.\ (1997) also discuss visually classified
morphologies on the revised Hubble scheme for the brighter galaxies
within this field down to $R_{702}=23.5$.  There are a total of 123
galaxies with morphologies from S97 within the joint {\it NICMOS}/{\it
WFPC2} field. All ten of the radio sources selected from the VLA map
are included in the S97 catalog and we list the Hubble type for each
galaxy in Table~2.  We also give the Disturbance Index ($D$), which is
a measure of how disturbed the galaxy seemed compared to the typical
appearence of a galaxy with its Hubble type locally, where $D=0$ is
``normal'' and $D=4$ is highly disturbed (S97).  In addition to these
objective measures, a subjective interpretation of the source of the
disturbance was also given by S97 and is listed in the table
(e.g.\ Chaotic, Tidal interaction, etc). We note in particular that a
galaxy would be classed as ``Merger'' if  two or more close nuclei were
seen in a common envelope.  These visual estimates were shown to
correlate well with the machine-based asymmetry measurements (S97) and
have been used to attempt to isolate and study the role of dynamical
interactions in triggering the Butcher-Oemler effect (e.g.\ D99). 

%
%
\hbox{~}\smallskip
\centerline{\psfig{file=f2.ps,angle=270,width=3.1in}}
\noindent{\scriptsize
\addtolength{\baselineskip}{-3pt} 
\hspace*{0.3cm}

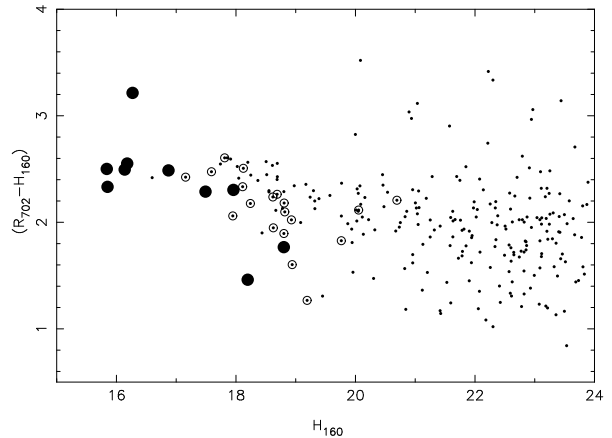
Fig.~2.\ The $(R_{702}-H_{160})$---$H_{160}$ color-magnitude
distribution for the sources lying in the joint {\it WFPC2}/{\it NICMOS}
field.  The $(R_{702}-H_{160})$ color is measured within an aperture of
diameter 2.5 times the FWHM of the galaxy in the F160W image.  The
large filled points denote those galaxies detected in the VLA map, the
circled points are other galaxies with spectroscopic identifications in
D99 and the small points are the remaining sources selected from the
F160W catalog.  Note the sequence of red galaxies associated with
early-type cluster members which extends down to $H_{160}\sim 23$.  

\addtolength{\baselineskip}{3pt}
}

To investigate the optical-infrared colors of the galaxies in our sample
we have aligned and resampled the {\it WFPC2} image to the reference frame
of the {\it NICMOS} exposure.  A comparison of stars and compact galaxies
in the resampled F702W exposure and the F160W image shows that they have
very similar profiles in the two bands indicating that the resolution
is dictated by the pixel sampling of the images.  This suggests that PSF
variations should not strongly influence our analysis.  To confirm this we
have convolved the {\it WFPC2} and {\it NICMOS} images with a model PSF for
the other instrument generated using {\sc tinytim} (Krist \& Hook 1999).
The colors determined from these convolved images show no systematic
differences from those measured from the unsmoothed images and hence we
use the latter in the following analysis.  

To measure colors from our F702W and F160W exposures we have used the {\sc
IRAF} task {\sc phot} and apertures matched to the near-infrared extent of
the galaxies.  The diameter of the photometry apertures were taken as 2.5
times the FWHM of each source from the F160W catalog ($d_{\rm phot}$ in
Table~2).  The color-magnitude diagram for the field is shown in Fig.~2,
with the radio sources and spectroscopically identified galaxies marked.
We give total $H_{160}$ magnitudes and aperture  $(R_{702}-H_{160})$
colors for the ten radio sources in Table~2.\footnote{For interest we list
here the $(R_{702}-H_{160})$ colors of the various $z>3$ lensed galaxies
identified by Trager et al.\ (1997).   These galaxies are all relatively
blue compared to the general field (Fig.~2), $(R_{702}-H_{160})=1.25\pm
0.10$ for DG\#334 at $z=3.34$, $(R_{702}-H_{160})=0.78\pm 0.15$ for the
$z=3.97$ source seen as P1/P2/P3 and $(R_{702}-H_{160})=1.43\pm 0.20$
for the $z=3.98$ arclet A0, although this may suffer contamination
from the nearby bright cluster galaxy (\#426, Fig.~3).  A search of the
{\it NICMOS} and {\it WFPC2} images to identify galaxies with extreme
colors turned up a very red disk galaxy near the center of the field:
\#333 (Fig.~1), this is the reddest, bright object in the field with
$H_{160}=19.53$ and $(R_{702}-H_{160})=5.4\pm 0.2$ (c.f.\ Treu et al.\
1999).} We show the F160W and F702W images of this sample in Fig.~3.

To study the color variations within the galaxies we have constructed
F702W$-$F160W color images for each galaxy in our spectroscopic sample.
We show the color images for the relevant radio-detected galaxies in
Fig.~3 to illustrate the internal color variations within these galaxies.
A comparison sample of confirmed cluster members which are not detected
in the radio map is shown in Fig.~4.  We note that PSF-convolved versions
of these images are qualitatively similar, in particular the blue cores
at the centers of \#296, \#399 and \#426 do not appear to result from
differences in the PSF between the two passbands.

%
%
\begin{table*}[hbt]
{\scriptsize
\begin{center}
\centerline{\sc Table 2}
\vspace{0.1cm}
\centerline{\sc Optical/Near-infrared Properties of the Radio Sample}
\vspace{0.3cm}
\begin{tabular}{cccccccccl}
\hline\hline
\noalign{\smallskip}
 {Source} & {$z$} & {$H_{160}$} & {($R_{702}-H_{160}$)} & $d_{\rm phot}$ & {Morph.} & {Spectral} & EW(H$\delta$)&  {$D$} & {Comments}\cr
 ID & & & & ($''$) & Type & Type & (\AA) &  &  \cr
\hline
\noalign{\smallskip}
~36 &  0.4119 &  18.20 & $1.46\pm 0.10$ & 5.6 & Sd &  e(a) &  10.2 & 2 & EW([O{\sc ii}])$=-6.9$\AA; Merger \cr
122 &  0.4125 & 17.96 & $2.30\pm 0.10$ & 2.7 & Sd &   a+k &  11.6 & 3 & Merger, H$\delta$ measurement uncertain. \cr
224 &  0.4076 & 15.85 & $2.33\pm 0.04$ & 5.8 & Scd? & k+a &  6.1 & 2 & Merger \cr
230 &   ...   & 18.80 & $1.77\pm 0.14$ & 2.2 & Irr &  ... &  ... & 2 & no spectrum in D99\cr
296 &  0.4014 & 16.87 & $2.49\pm 0.07$ & 2.5 & E   & k+a   &  4.6 & 0 &\cr
399 &  0.4109 & 16.14 & $2.50\pm 0.05$ & 2.5 & E   & k     &  0.0 & 0 & \cr
426 &  0.4037 & 16.18 & $2.55\pm 0.05$ & 3.9 & E   & k    &  0.0 & 0 & \cr
536 &  0.4111 & 17.49 & $2.29\pm 0.08$ & 4.0 & Sc & a+k   &  8.6 & 3 & Tidal? \cr
610 &  0.4007 & 15.84 & $2.50\pm 0.05$ & 3.4 & Sa/S0 & k+a & 4.2 & 1 & Merger?  \cr
640 &  0.3324 & 16.27 & $3.21\pm 0.04$ & 8.3 & Sab  & k:   & 0.0  & 1 & Field galaxy, low s/n spectrum --- possible e(a)?\cr
\noalign{\hrule}
\noalign{\smallskip}
\end{tabular}
\end{center}
}
\end{table*}

\subsection{Spectral Information}

Spectroscopic observations of galaxies within our field are available
from the MORPHS spectral catalog (D99).   The targets for these
observations were selected from the F702W exposure and D99 amassed
spectra for 28 galaxies within the {\it NICMOS} field, comprising 22
cluster members and 6 field galaxies.  These spectra typically cover the
spectral region from 3000--6000\AA\ in the restframe, including any
[O{\sc ii}]$\lambda$3727 emission but in most cases not H\,$\alpha$.  The
galaxies have all been classified using the spectral classification scheme of
D99 as follows: passive galaxies (k); post-starburst (k+a/a+k), these
show no [O{\sc ii}] and strong Balmer absorption indicative of a
cessation of star formation within the last Gyr or so; actively star
forming (e(c)/e(b)/e(a)), e(c) have spectral properties similar to
local spiral galaxies, e(b) have stronger [O{\sc ii}] and are
interpreted as starburst galaxies, e(a) have moderate [O{\sc ii}] but
stronger Balmer lines than e(c) galaxies.  The e(a) class has been
interpreted as dusty starbursts (P99) where the [O{\sc ii}] is
suppressed by dust within the galaxy (supported by the observations of
anomalously high H$\alpha$/[O{\sc ii}] ratios for these galaxies, see
also \S1).  The distribution of cluster/field galaxies in the different
spectroscopic classes is:  k (7/2); k+a (6/0); a+k (2/0); e(a) (5/1);
e(b) (1/0); e(c) (2/1).  The spectrum of one field galaxy has too low
signal-to-noise to allow a reliable classification.  

For the ten radio sources in the field, spectra and spectral
classifications are available from D99 for nine galaxies (galaxy \#230
lacks a spectroscopic ID) and we list these in Table~2.  Of these nine
galaxies, eight are cluster members and one is a field galaxy (the Sab
galaxy \#640).

\section{Results}

With a small sample of galaxies our conclusions are by necessity fairly
qualitative in nature and this paper should be seen as the prelude to
the analysis of a larger sample of radio-selected galaxies in distant
clusters by Morrison et al.\ (1999).  Nevertheless, we will use our
observations, and particularly the high resolution near-infrared
imaging with {\it NICMOS}, to illustrate the general characteristics of
radio galaxies in distant clusters and to emphasise the effects of dust
in galaxies on their perceived properties.

\subsection{Radio-selected Sample}   

We have detected ten galaxies above the 3$\sigma$ flux limit
($S_{1.4}\geq 27\mu$Jy) of the VLA map which lie within the joint {\it
NICMOS}/{\it WFPC2} field.  The spectroscopic information available
from D99 allows us to identify eight of these galaxies as cluster
members (or 35\% of the 22 cluster galaxies known within this field)
and a further source as a foreground field galaxy (12\% of the
spectroscopically-confirmed field population within our image).  Our
field size corresponds to a region $\sim 650\times 650$\,kpc in extent
at the cluster redshift and the equivalent radio power limit is
$2\times 10^{22}$\,W\,Hz$^{-1}$.  A comparably deep survey of a similar
region in the core of a rich local cluster would typically detect $\ls
1$ galaxy (e.g.\ Fig.~5), compared to the eight detected here.  This
suggests a substantial increase in the rate of occurence of radio
sources in high density environments at moderate redshifts (Morrison et
al.\ 1999).

The galaxies included in the radio-selected sample of cluster members
are typically the brighter galaxies in our field (Fig.~2), including
the brightest four galaxies (in the restframe $B$-band). The whole
sample spans a range of magnitudes $R_{702}=18.3$--20.8 ($L_B \sim
4$--$0.4L_B^\ast$).  The colors of the fainter radio-detected galaxies
are slightly bluer than equivalently luminous, undetected cluster
members, with the brighter radio-emitting galaxies having comparable
(red) colors to the undetected population (Fig.~2).  The fact that we
only detect the optically brightest galaxies in the radio map may
suggest that the majority of cluster galaxies would be detected if we
could increase the sensitivity of the radio observations by a factor of
a few (see Fig.~5).\footnote{We note that the radio population has
similar dynamics to those cluster members undetected in the radio map: we
calculate a restframe velocity disperion of $\sigma = (960\pm
160)$\,km\,s$^{-1}$ for the eight radio-detected members compared to
$\sigma = (1110\pm 140)$\,km\,s$^{-1}$ for the 12 undetected members
lying within the same field.} 

In our analysis we will assume that the local relationship between star
formation rate and radio emission (e.g.\ Condon 1992) holds for spiral
galaxies detected in distant clusters.  The validity of the local
relationship at least for the distant field appears to be supported by
recent observations (Georgakakis et al.\ 1999).   However, we caution
that the exact star formation rates derived from this analysis may have
substantial systematic uncertainties.  In particular the claim that the
radio to far-infrared correlation varies locally between galaxies in
rich clusters and those in the field (Andersen \& Owen 1995) may
indicate an environmental influence in the conversion from radio
luminosity to star formation rate (SFR) resulting from the compression
of the galactic magnetic field by the galaxy's motion through the
intracluster medium (ICM).  Without a fuller understanding of the state
of the ICM in distant clusters it is difficult to quantify the extent
of this bias on the radio-derived SFRs of distant cluster galaxies. We
have chosen therefore to quote the SFRs determined using the local field
relationship and the correction which should be applied if the radio
population in distant clusters shows the same behaviour as is claimed
for galaxies in rich clusters locally.

\subsection{Optical Morphologies}

Turning to the restframe $B$-band morphologies of the radio sources
(Table~2 \& Fig.~3) we see that they fall roughly equally into two broad
classes:  optically-luminous early-type galaxies (E and S0/Sa) and mid-
to late-type spiral galaxies (Sc--Sd/Irr).    

In their analysis of galaxies in the Virgo cluster detected in the NVSS
radio survey, Gavazzi \& Boselli (1999) found that the most powerful
radio sources in spirals were also in mid-type, Sb--Sc, systems. To
compare the optical and radio properties of the galaxies we plot the
1.4-GHz restframe power versus the optical $B$-band luminosity for all
spectroscopically-confirmed members in Fig.~5, separated into early-
and late-type.  The early-type galaxies show a wider range in their
$P_{1.4}/M_{B}$ values than the spirals, but span a similar region of
the $P_{1.4}$--$M_{B}$ plane as radio-selected E/S0 galaxies in the
Virgo cluster (Fig.~5).  This is as expected if the radio emission from
the early-type galaxies is coming from a central AGN. In contrast the
radio emission from the spirals is (by analogy to local systems) most
likely synchotron emission associated with active star formation and
leading to a correlation between the radio power and the optical
luminosity of the galaxy.  We note that the population of
radio-emitting spiral galaxies identified by Gavazzi \& Boselli (1999)
in  Virgo  have radio powers $5\times$ lower at a fixed $B$-band
luminosity than the galaxies we are detecting in Cl\,0939+4713 (Fig.~5).

%
%
\hbox{~}\smallskip
\centerline{\psfig{file=f5.ps,angle=0,width=3.1in}}
\noindent{\scriptsize
\addtolength{\baselineskip}{-3pt} 
\hspace*{0.3cm}

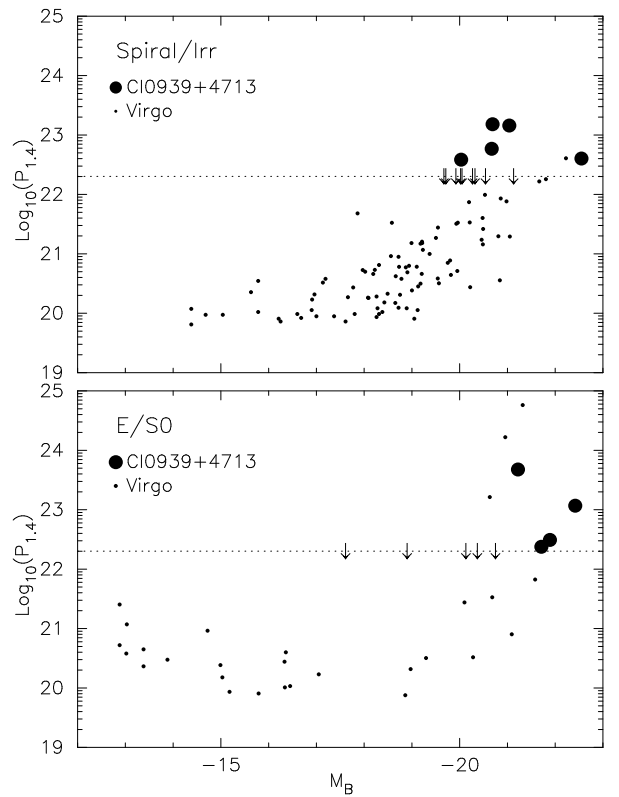
Fig.~5.\ The $P_{1.4}$--$M_B$ distribution for those galaxies detected in
the VLA map of Cl\,0939+4713 compared to the equivalent distribution for
the Virgo cluster (Gavazzi \& Boselli 1999).  The top panel compares the
distributions for the morphologically-classified spiral and irregular
galaxies in both clusters, while the lower panel shows the equivalent
comparison for elliptical and S0 members.  We adopt an apparent
magnitude of $R_{702}=19.7$ for an $M_B=-21.1$ ($L_B^\ast$) galaxy in
the $R_{702}$-band observed at $z=0.41$ (assuming an SED typical of a
present-day Sbc galaxy).  The dotted lines show the radio flux limit of
our sample in Cl\,0939+4713.  We indicate the spectroscopically-confirmed
member galaxies which are undetected in the VLA map as upper limits.

\addtolength{\baselineskip}{3pt}
}

The association of around 60\% of the bright radio-selected sample with
spiral galaxies suggests that the  evolution in the radio populations
within clusters runs in parallel to the increasing fraction of blue
star-forming galaxies found in these regions at $z\gs 0.2$, the
Butcher-Oemler effect. The rise in the fraction of bright spiral
galaxies in clusters such as Cl\,0939+4713 is a factor of $\sim 5$ out
to $z=0.4$ (Dressler et al.\ 1997).  Even without the increased radio
luminosity of the brightest cluster spiral galaxies, this increase in
the proportion of spirals is enough to explain the evolution in the
number of radio sources seen in the distant clusters (a factor of $\sim
5$--6, Morrison et al.\ 1999).  The increase in the spiral fraction in
the clusters is believed to result from infall from the surrounding
field associated with the growth of the clusters.   A rise in the radio
activity of moderate redshift field galaxies has also been observed and
attributed to star formation in these galaxies (Georgakakis et
al.\ 1999).  However, to fully understand the evolution of the radio
population in the cluster we also need to identify the cause of the
increased radio luminosities of the bright spiral galaxies.  We note
that our use of a local cluster sample in the comparison should
minimise the possible environmental effects which could explain this
apparent brightening, and moreover the increase in radio luminosities
is substantially above that claimed to result from environmental
effects within local clusters (\S3.1).

\subsection{Near-infrared Morphologies}

We now discuss the internal structure of the cluster galaxies in our
{\it HST} images (Fig.~3 and 4) and the important insights which our
{\it NICMOS} near-infrared imaging has provided into the nature of
these systems.  Starting with the early-type galaxies we see relatively
uniform optical-near-infrared colors (as shown by their lack of
structure in the F702W$-$F160W images in Fig.~3 and 4).  A few contain
blue cores, which may be associated with central AGN, although the
details of these features are sensitive to the exact form of the PSF in
the F702W and F160W images and so we do not discuss them further here.

A comparison of the F702W and F160W morphologies for the late-type
radio sources in Fig.~3 shows that in these systems there are
substantial differences in the appearence of these galaxies between the
restframe $B$- and $J$-bands. The near-infrared morphologies are
considerably more relaxed and symmetrical.   However, the lack of a
morphological classification scheme for local galaxies based on
near-infrared imaging means that is not useful to attempt to reclassify
the galaxies on the basis of the F160W images. 

More interestingly, the F702W$-$F160W images in Fig.~3 show that in a
large number of cases the irregular structures seen in the rest-frame
$B$-band within the central regions of the late-type galaxies are
replaced by a much more regular appearence in the near-infrared and so
must be due to very red structures (visible as white regions in
Fig.~3).  These features are substantially redder than the outskirts of
the galaxies (up to $\Delta (R-H) \sim 1.3$--1.5) and could represent
differences in the stellar populations within the galaxies or simply
dust obscuration.  To distinguish between these alternatives we
concentrate on the large fraction of galaxies whose integrated spectra
show post-starburst stellar populations.  The lack of emission lines in
the optical spectra indicates that there is no visible star-formation
in the outskirts of these galaxies, which also show the red
optical-infrared colors typical of an evolved stellar population (see
the integrated colors in Fig.~2).  In this situation it is very
difficult to make the stellar population in the central regions of the
galaxies substantially redder than the outskirts without invoking a
contrived scenario. In contrast, dust obscuration can easily produce
both the apparently red colors and its structured distribution.  We
propose therefore that large quantities of dust are probably a common
feature of these galaxies.

The small scale features seen in the optical images of these galaxies
which we claim are due to dust obscuration have been previously
interpreted as arising from dynamical disturbance of these galaxies,
suggesting a dynamical origin for their star formation activity (S97).
To understand the consequences of this change in interpretation of these
galaxies we focus on those galaxies classified in the optical as showing
signs of morphological disturbance by S97. There are   a total of eight
spectroscopically confirmed members in our sample with  disturbance
indices of $D\geq 2$ indicating abnormally disturbed morphologies, all
of these are disk galaxies and they are typically classified as mergers
or tidally disturbed, half of them are detected in our radio map.
These galaxies exhibit substantially more regular morphologies in
the near-infrared (Figs.~3 \& 4), in particular the apparently double
structures seen in the F702W morphologies of the radio sources \#36,
\#122 and \#536 all disappear in the F160W passband. 

We also see signs of dust obscuration in the disturbed galaxies which are
not radio sources (\#171, \#369, \#497 and \#622), although arguably not
as extreme as the radio-selected sample.  In a few cases there are other
signatures of interactions in the F702W images apart from double nuclei,
etc, such as low-surface brightness tidal features or companion galaxies
(e.g.\ \#224 or \#497), which indicate that these are true merging or
interacting systems. But, in the majority of the galaxies their high
disturbance index was based upon the apparent presence of multiple nuclei
in the central regions of the galaxies (see \S1).  We suggest that in fact
dust obscuration is the correct explanation for their irregular appearence
in the restframe $B$-band which results in their  classification as
disturbed or merging.  We estimate that of the eight confirmed members
with $D\geq 2$ based on the F702W imaging, at least half (\#36, \#122,
\#171, \#622) would have $D<2$ if classified from the F160W images,
suggesting that the occurence of highly disturbed galaxies in the core
of Cl\,0939+4713 may have been overestimated by up to a factor of two.

We conclude from our comparison of the near-infrared and optical
morphologies of a small sample of confirmed cluster members that the
frequency of galaxies showing small-scale signatures of disturbance or
interaction is probably substantially less than has been suggested from
optical studies (e.g.\ S97) due to obscuration by dust.  This effect
acts in addition to the standard luminosity bias in optically-selected
samples towards actively star-forming galaxies (the usual justification
given for undertaking near-infrared surveys of distant clusters,
e.g.\ Barger et al.\ 1996).  The expectation of previous
optically-based studies  was that dust obscuration would not be a
significant bias when comparing the distant morphological samples with
more local examples as they were comparing effective restframe $B$-band
observations in both cases (S97).  However, this expectation seems to
have been misleading due to the wider prevelance of dust in cluster
galaxies at $z\sim 0.4$, perhaps associated with the increased activity
in galaxies at these epochs.

\subsection{Spectral Properties}

Looking now at the spectral properties of the radio-selected cluster
members we see something striking:  five of the galaxies have spectral
features which class them as post-starburst (a+k/k+a) systems using the
scheme of P99 (Table~2). Moreover, these five radio sources show
the strongest Balmer absorption lines of the eight post-starburst
galaxies in our cluster sample (they are also the brightest five
post-starburst galaxies). There are no post-starburst field galaxies in
our sample.
  
The detection of post-starburst galaxies in the radio waveband is
surprising. These galaxies are selected to have no detectable [O{\sc
ii}] emission and hence are expected to have no on-going massive star
formation (observations of the spectral region including any H$\alpha$
emission in these galaxies would make this statement more robust).  In
contrast, the radio emission is most easily explained as arising from
massive star formation (although some contribution from an obscured
radio-loud AGN cannot be ruled out with our present data).  The
expected lifetime of the radio emission after the starburst is $\ll
0.1$\,Gyr, thus if they are post-starburst systems these galaxies must
all be seen {\it just} after the cessation of their star formation.
This seems very unlikely as the length of time in which the
post-starburst signature (strong H$\delta$ absorption) is visible is
$\sim 1$\,Gyr and we would thus expect to see an order of magnitude
more radio-quiet k+a galaxies, where the radio emission had decayed but
the Balmer absorption is still visible.  This population is not
observed.  Therefore, we conclude that massive stars are currently
being formed in these galaxies.

Where is the site of the active star formation in the a+k/k+a cluster
galaxies?  The optical spectra typically sample the whole of the bulge
and disk of the galaxy (see D99) and thus it is unlikely that we are
missing emission from the outskirts of the galaxy.  We suggest instead
that the star formation is occuring in the central regions of these
galaxies but is hidden from view by the dust which we see there in our
F702W$-$F160W images.  This possibility was raised in P99 who stated
that the most extreme e(a) galaxies might appear as post-starburst
systems (a+k/k+a) due to obscuration of the emission regions.
The reddened regions lie in the central 5--10\,kpc of the galaxies
(Fig.~3 \& 4) and have optical--near-infrared colors which are typically
$\Delta(R_{702}-H_{160})\sim 1$ magnitude redder than the outskirts
of the galaxies.  This would indicate optical extinctions of at least
$A_B\sim 2$\,mag to the population of disk stars (let alone the active
sites of star formation) although this estimate is highly uncertain due
to resolution effects.  

Four of the five radio-selected post-starburst galaxies are disk
galaxies and we can estimate their star formation rates (SFR) from
their radio power using the calibration of Condon (1992). We find that
these galaxies are typically forming massive stars ($M\geq 5M_\odot$)
at rates of $\sim 10$\,$M_\odot$ yr$^{-1}$, with the highest SFRs being
30\,$M_\odot$ yr$^{-1}$ for \#122 and \#536. This assumes no boosting of
the radio emission due to compression of the galactic magnetic fields by
the cluster ICM, which could amount to a factor of $\sim 2$ (Andersen \&
Owen 1995).  Combining these SFRs with the $B$-band luminosities of the
galaxies we can predict their  [O{\sc ii}] fluxes if the star formation
regions are unobscured using the relations given in Kennicutt (1992).
We can then roughly estimate the amount of extinction needed for this
line to be undetected in the optical spectra, EW([O{\sc ii}])$\ls 5$\AA.
We find that the star forming regions would need to suffer $A_B \gs
3$\,mag of extinction for the [O{\sc ii}] line to be undetected in
the D99 spectra ($A_B \gs 2$\,mag if we assume the radio fluxes are
artificially raised).  This amount of obscuration is not unreasonable
given the extent of the dust reddening seen in the F702$-$F160W images
(Figs.~3 \& 4).  We note that if we assume this dust is cold, $T_D \sim
40$\,K, then the predicted sub-mm fluxes of these galaxies will be only
$S_{850}\sim 0.5$\,mJy at 850\,$\mu$m.  This is well below the confusion
limit of current sub-mm instrumentation in blank fields.

There are also examples of proposed dusty galaxies within our sample
which are not detected in the radio map. In particular three of the
cluster galaxies have e(a) spectral features and evidence for dust in
their F702$-$F160W images (Fig.~4, \#177, \#369 and \#622).  These
galaxies are optically fainter than the e(a) galaxy which is detected
in the radio, \#36, and it is possible that their lower overall
luminosities mean that their radio emission is below the limit of the
VLA map.  Interestingly, although the e(a) spectral class was
interpreted as dusty starbursts in P99, the radio map places upper
limits on the current rate of massive star formation in the undetected
e(a) galaxies of $\Mdot \ls 5 M_\odot$ yr$^{-1}$, with similar limits
applying to the galaxies with e(c) spectra in the field (e.g.\ \#165
and \#497 in Fig.~4).  Taking the possible boosting of the radio
emission due to compression of the galactic magnetic fields by the
cluster ICM (\S3.1) into account these SFR limits could be dropped by a
factor of two or more.

\section{Discussion}

We see a combination of post-starburst spectral features, radio
emission and apparently large quantities of dust within the spiral
galaxies in our sample.  These characteristics can be most easily
accomodated if they exist in spatially distinct regions of the
galaxies, as suggested by our color maps. The radio emission would
arise from massive star formation in the dust obscured center of the
galaxy, while the A stars trace the remanents of an earlier ($\ls
1$\,Gyr) strong star formation episode within the galactic disk.  There
is weak evidence in our sample for a correlation between H$\delta$ line
strength and radio luminosity which would suggest that there is a
causal link between the numbers of A stars and the current (obscured)
star formation rate, probably as a result of a single triggering
mechanism.

This triggering mechanism could be a cluster-related phenomenum, a view
which is supported by the lack of obvious post-starburst systems in
samples of faint radio sources in the field (Hammer et al.\ 1995;
L.L.\ Cowie, priv.\ comm.). Thus, we may be seeing a more dynamic
situation including cluster-driven processes, whereby interactions
between the intracluster medium and the cold molecular gas within the
galactic disks accelerates the collapse and evaporation of the star
forming regions.  The effects of the ICM on the galactic disk are
expected to be a function of the local density in the disk -- with
lower density regions in the outskirts of the disks being most severely
effected, while star formation continues almost uneffected in the inner
regions of the galaxy.  Thus environmental processes could  result in a
combination of post-starburst and starburst within individual cluster
galaxies.  Studies of the spatial distribution and kinematics of larger
samples of radio-selected galaxies may be able to discriminate between
the different physical processes which could be operating (Morrison et
al.\ 1999).

Alternatively, perhaps this activity may be triggered by a more general
physical process, unrelated to the cluster environment.  Indeed,
similar behaviour is seen in the models of interacting field galaxies
published by Mihos \& Hernquist (1996) (although we have argued that a
substantial fraction of apparently disturbed cluster spiral galaxies
probably arise from dust obscuration, rather than true interactions).
For a range of models for tidal interactions between galaxies Mihos \&
Hernquist predict that spatially-extended enhanced star formation will
first occur in the disk of the galaxy.  This is followed on a timescale
of a $\sim 0.1$--0.3\,Gyr by funnelling of any remaining gas into the
center of the galaxy where it reaches very high densities and triggers
an (obscured?) nuclear starburst.   Again, more detailed investigation
of the kinematics and internal dynamics of these galaxies may provide
useful diagnostics to distinguish between the competing models. 

The picture which is appearing is one in which galaxies which are
visibly forming stars appear to have modest star formation rates, $\ls
5 M_\odot$ yr$^{-1}$ ($M\geq 5M_\odot$), while more vigorous star
formation, $\gs 10$--$30M_\odot$ yr$^{-1}$ is on-going in highly
obscured regions in the centers of what appear to be post-starburst
galaxies.  If the radio fluxes of these distant galaxies are increased
by the same environmental effects which are claimed to boost the radio
emission of local cluster galaxies and to the same extent (Andersen \&
Owen 1995) then the quoted SFRs should be reduced by a factor of $\sim
2$.  Sensitive observations to search for H$\alpha$ emission (or
emission lines further into the near-infrared) will provide a stronger
test of the star formation rates in these galaxies, the extent of
obscuration within them and the degree to which their radio emission deviates
from that seen in the distant field population due to environmental
processes. Finally we note that by having post-starburst and starburst
phases occuring in parallel within a single galaxy it is possible to
reduce the  high fraction of galaxies ($\sim 100$\%, Barger et
al.\ 1996) which must pass through this phase to explain the large
numbers of post-starburst systems seen within distant clusters (Barger
et al.\ 1996; P99).

\section{Conclusions}

In summary, we have highlighted the role that dust plays in obscuring
our view of the morphologies and star formation properties of distant
galaxies.  We have shown that selection at faint radio fluxes provides a
powerful technique for identifying spectrally-classified post-starburst
and starburst galaxies crucial to our understanding of the evolutionary
cycle of galaxies in clusters (e.g.\ P99). Using this new tool we can
undertake statistical studies to search for the mechanism responsible for
triggering and/or quenching of this activity (P99; Morrison et al.\ 1999).
The main conclusions of our study are:

\noindent{$\bullet$} We present optical and near-infrared imaging
with {\it HST} of a region in the core of the distant rich cluster
Cl\,0939+4713 at $z=0.41$.  

\noindent{$\bullet$} We compare the high resolution near-infrared and
optical morphologies of a sample of 22 cluster members. This shows that
while the broad morphological classifications are similar, the
frequency of galaxies showing small-scale signatures of disturbance or
interaction is substantially less when using the near-infrared
imaging.  We suggest that obscuration by dust has led optical studies
to conclude that an anomalously high fraction of galaxies have suffered
small scale disturbances. This may point to a wider prevelance of dust
in cluster galaxies at $z\sim 0.4$, perhaps as a result of the
increased star formation activity at these epochs.

\noindent{$\bullet$} Using a very deep VLA 1.4-GHz map of this area we
select a sample of ten radio galaxies in our field above a flux limit
of $S_{1.4}=27\mu$Jy, equivalent to a restframe 1.4-GHz power of $2
\times 10^{22}$ W\,Hz at the cluster redshift.  The faint radio sources
are associated with two populations of cluster galaxies: luminous
early-type galaxies, the radio emission from these galaxies originates
from AGN; and dusty galaxies with late-type spiral morphologies, the
radio emission here is expected to trace massive star formation
(although the exact relationship between the radio luminosity and the
SFR is uncertain due to environmental influences).  

\noindent{$\bullet$} The optical spectral characteristics of the
late-type radio members are highly unusual -- they appear to be
post-starburst systems.  We suggest that these galaxies are not
post-starburst but in fact host highly obscured star-formation and
starbursts.   This is supported by our high resolution
optical--near-infrared imaging which shows that the central regions of
these galaxies are heavily dust enshrouded.  If they continued for
several $100$\,Myrs, these dust enshrouded nuclear-starbursts could
build a substantial stellar bulge in the remanent of these luminous
disk galaxies, possibly contributing to the formation of the S0
population in the clusters (Mihos \& Hernquist 1996).

\noindent{$\bullet$} We emphasise that a lack of consideration of the
effects of dust on the morphological and spectral properties of galaxies,
even at the relatively modest redshifts discussed here, may lead to
an incomplete understanding of the cycle of star formation in cluster
galaxies.  Further work on this subject is urgently required.

\noindent{$\bullet$} This paper is the prelude to a larger study of the
{\it HST} morphologies and spectral properties of radio-selected galaxies
in Cl\,0939+4713 and other high redshift clusters (Morrison et al.\ 1999).
We expect that extensive studies with higher spatial resolution and wider
wavelength coverage in the near future using the upgraded VLA and the
next generation of millimeter interferometers will provide new insights
into the prevelance and distribution of dust in distant galaxies.

\section*{Acknowledgements}

This paper is based upon observations obtained with the NASA/ESA {\it
Hubble Space Telescope} which is operated by STSCI for the Association
of Universities for Research in Astronomy, Inc., under NASA contract
NAS5-26555 and on observations from the Very Large Array which is
operated by Associated Universities, Inc., under a cooperative
agreement with the National Science Foundation.  We thank Lisa
Storrie-Lombardi and Lin Yan for help with {\it NICMOS} photometry and
Giuseppe Gavazzi and Alessandro Boselli for kindly providing the data
on radio galaxies in Virgo in electronic form.  We also thank John Hill
and the {\it NICMOS} GTO team for allowing us to observe this target.
We acknowledge the use of the Pedestal Estimation and Quadrant
Equalization Software developed by Roeland P.\ van der Marel.  We thank
an anonymous referee for detailed comments on this work and Amy Barger,
Andrew Blain, Warrick Couch, Len Cowie, Alan Dressler, Alastair Edge,
Gus Oemler and especially Bianca Poggianti for useful conversations.
IRS acknowledges support from the Royal Society and RJI from PPARC.

\clearpage

%
%
\begin{figure*}
\noindent{\scriptsize
\addtolength{\baselineskip}{-3pt} 
\hspace*{0.3cm}

Fig.~1.\ The VLA 1.4-GHz map overlayed as a contour plot on the {\it
HST} {\it NICMOS} F160W mosaic.  The radio sources are identified in
large font using the numbering scheme of Smail et al.\ (1997).  The
galaxies labelled with the smaller font represent other
spectroscopically identified galaxies (with the exception of \#333)
from the catalog of Dressler et al.\ (1999), these are similarly
numbered using the S97 identifications.   The dashed line shows the
overlap on this field of the {\it WFPC2} F702W exposure from S97. The
radio contours are at 12, 25, 50, 100 and 200\,$\mu$Jy per beam.  
North is top and east is left in this figure.

\addtolength{\baselineskip}{3pt}
}
\end{figure*}

%
%
\begin{figure*}
\noindent{\scriptsize
\addtolength{\baselineskip}{-3pt} 
\hspace*{0.3cm}

Fig.~3.\ The groups of {\it NICMOS} F160W, F702W$-$F160W and {\it
WFPC2} F702W images for each of the ten galaxies selected from the VLA
1.4-GHz map of Cl\,0939+4713.  The F160W and F702W images provide
restframe $J$- and $B$-band views of the galaxies and these are shown
with a linear intensity scale, while the color image is constructed
from the difference of the $\log$-scaled F160W and F702W images.  In
the F702W$-$F160W color image lighter regions represent redder areas of
the galaxy, the typical color range is roughly $R_{702}-H_{160}=2.4$--3.2 
(black to white). The panels are labelled with the catalog number,
morphological type, spectral type and disturbance index (from S97 and
D99).  Each panel is $10''\times 10''$ (equivalent to 65\,kpc square at
the cluster redshift) with north top and east left and the panels are
ordered on morphology.  Note that \#640 lies in the extreme corner of
the {\it WFPC2} field and \#230 falls on a chip-boundary in the F702W
exposure.  All of the galaxies are confirmed cluster members with the
exception of \#640 which is a foreground field galaxy and \#230 which
has no spectroscopic identification.
   
\addtolength{\baselineskip}{3pt}
}
\end{figure*}

%
%
\begin{figure*}
\noindent{\scriptsize
\addtolength{\baselineskip}{-3pt} 
\hspace*{0.3cm}

Fig.~4.\ A comparison sample of spectroscopically-confirmed cluster
galaxies which show no detectable radio emission in the VLA map above
a 1.4-GHz flux of $S_{1.4}=27\mu$Jy.  This figure includes all the
spectrally-active members of this sample and again is ordered on
morphology.  The panels are labelled with the galaxy ID, morphology
and disturbance index, D, from S97 and the spectral classification from
D99.
   
\addtolength{\baselineskip}{3pt}
}
\end{figure*}

\end{document}